\documentclass[11pt,twocolumn]{article}
\usepackage[utf8]{inputenc}
\usepackage[T1]{fontenc}
\usepackage{lmodern}
\usepackage[a4paper, top=20mm, bottom=43mm, left=19mm, right=19mm]{geometry}
\usepackage{amsmath,amssymb}
\usepackage{graphicx}

\usepackage[numbers,sort&compress]{natbib} 
\usepackage[colorlinks=true,
            linkcolor=blue,
            citecolor=blue,
            urlcolor=blue]{hyperref}

\usepackage{sectsty}
\sectionfont{\large\bfseries}      
\subsectionfont{\normalsize\bfseries}

\usepackage{epstopdf}
\usepackage{url}
\usepackage[showboxes]{textpos}
\usepackage{siunitx}
\usepackage [lined, algoruled]{algorithm2e}
\usepackage{algpseudocode}
\usepackage{authblk}
\usepackage{graphicx}
\usepackage{epstopdf}
\usepackage[numbers]{natbib}
\usepackage[font=small,labelfont=bf]{caption}
\usepackage{tabularx}  
\usepackage{booktabs}
\usepackage{amsthm}

\newtheorem{definition}{Definition}

\usepackage{abstract}
\renewenvironment{abstract}{%
  \small
  \begin{center}%
    \bfseries Abstract
  \end{center}
}

\usepackage{fancyhdr}
\fancypagestyle{firstpage}{%
  \fancyhf{}
  \fancyfoot[L]{%
    \begin{minipage}[t]{\textwidth}%
      \vspace{2pt}
      \scriptsize
      \noindent
      © 2016 Association for Computing Machinery. This is the author’s version of the work. 
        It is posted here for your personal use. Not for redistribution. 
        The definitive Version of Record was published in ACM Transactions on Mathematical Software, Vol. 43, No. 3, Article 18, pp. 1–17, August 2016, \url{https://doi.org/10.1145/2935745}
        
    \end{minipage}%
  }%
}

\begin{document}

\title{Systematic Alias Sampling: an efficient and low-variance way to sample from a discrete distribution}

\author{ Ilari Vallivaara,
Katja Poikselk\"a,
Pauli Rikula,
 Juha R\"oning
\affil{Department of Computer Science and Engineering (CSE), University of Oulu}
}
\date{2016}

\maketitle

\thispagestyle{firstpage}

\begin{abstract} In this paper we combine the Alias method with the concept of systematic sampling, a method commonly used in particle filters for efficient low-variance resampling. The proposed method allows very fast sampling from a discrete distribution: drawing $k$ samples is up to an order of magnitude faster than binary search from the cumulative distribution function (cdf) or inversion methods used in many libraries. The produced empirical distribution function is evaluated using a modified Cram{\'e}r-Von Mises goodness-of-fit statistic, showing that the method compares very favourably to multinomial sampling. As continuous distributions can often be approximated with discrete ones, the proposed method can be used as a very general way to efficiently produce random samples for particle filter proposal distributions, e.g. for motion models in robotics. 
\end{abstract}

\section{Introduction}
\begin{figure} [ht]
 \centering

 \includegraphics[width = \columnwidth]{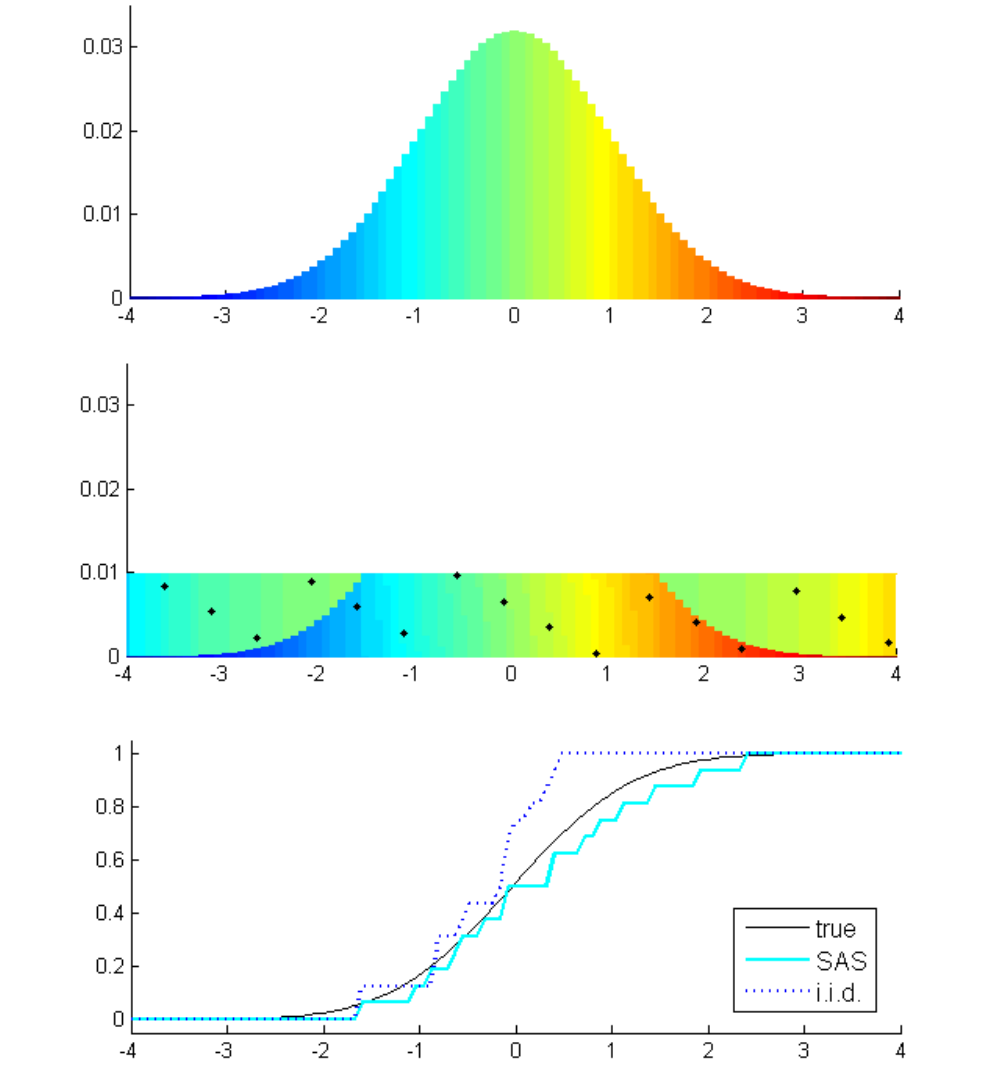}
 \caption{ Top: Discrete 101-valued approximation of the standard normal distribution with support of $[-4,4]$ denoted as $\hat{N}_{101}$. Middle: Alias table structure generated from $\hat{N}_{101}$. The probabilities of selecting the aliased values are depicted as the upper part of the bins. The colours correspond to values in the topmost figure. The black dots are an example batch of 16 systematic samples. Bottom: The empirical distributions defined by the 16 systematic (alias) samples (SAS) and 16 i.i.d. samples compared to the true cumulative distribution function. }
 \label{fig:aliasTable}
\end{figure}

There are many ways to generate random samples from an arbitrary distribution. Most of them are based on generating a random number in the unit interval and then using some method to map the number on desired distribution \cite{niederreiter92random}. One example for discrete distributions, with $n$ finite but potentially a large amount of possible values, is to generate the samples in $O(\log n)$ by using a binary search on the cumulative distribution function (cdf). Better performance can be achieved  using the Alias method introduced by Walker \cite{walker74new, walker77efficient}, that allows sampling in constant time. The method is based on constructing an alias table of the distribution, where each equiprobable bin contains at most two different values (see Figure \ref{fig:aliasTable} for illustration). The construction can be done in linear time \cite{kronmal79alias}. When sampling, one first determines the bin and then randomly selects the final value according to value proportions in the bin. For example, because in robotics the motion models are typically time-invariant \cite{thrun2001robust}, the alias tables for the needed distributions can usually be precomputed.

When we have to create $k$ samples at a time, we can improve the time performance by allowing the samples in the batch to be mutually dependent. One example of this is systematic resampling \cite{douc05comparison, hol2006resampling, arulampalam02tutorial} used widely in particle filtering, where one draws $k$ samples from a discrete distribution ($n = k$) corresponding to the particle weights. Despite having the same time complexity $O(n + k)$ with efficient implementation of independent and identically distributed (i.i.d.) roulette wheel sampling, this method firstly avoids the creation of $k-1$ random numbers and secondly produces samples with reduced sampling variance. Although it is possible to construct pathological distributions not suitable for systematic sampling, the method is usually considered preferable to i.i.d. resampling. However, when sampling from distributions where the number of possible values $n$ is much larger than $k$, the effectiveness of systematic sampling can be questioned. In this paper we present a systematic method aiming for efficiency regardless of $n$ without sacrificing much of the low-variance properties of systematic sampling.

\section{Motivation}
Drawing samples repetitively from a distribution is widely required in Monte Carlo methods, such as particle filtering commonly used in robotics \cite{thrun2001robust, thrun05probabilistic}. Many mathematics libraries, such as Apache Commons Math (ACM) for Java \cite{apache33}, include quite effective sampling methods for selected common distributions. For example, the ACM NormalDistribution \textit{sample}() method  uses internally by default the Well19937c random number generator's (RNG) \textit{nextGaussian}() method that is implemented as the Box-Muller transform. Generating normally distributed samples takes only roughly 2.5 times as long as the Well19937c \textit{nextDouble}() that is used internally by the Box-Muller transform. However, being able to sample from an arbitrary distribution is beneficial, as non-Gaussian approximations for proposal distributions often lead to more accurate results \cite{stachniss2007analyzing}. Sampling from an arbitrary distribution typically relies either on rejection sampling or binary search on the discretized cdf of the distribution, for which we propose an effective and low-variance alternative in this paper. 

In particle filtering and robotics literature systematic resampling is well studied \cite{douc05comparison, hol2006resampling, arulampalam02tutorial, thrun05probabilistic}. However, when generating samples for proposal distributions, e.g. for motion models, the sample generation details are not usually mentioned, with few exceptions \cite{ormoneit2001lattice}. From this we assume not much attention is generally paid to sampling efficiency outside the resampling step. However, in our research \cite{vallivaara13monty, vallivaara11magnetic} we felt that being able to easily construct arbitrary distributions and sample effectively and conveniently from them would be very beneficial. 

The need for time efficiency arises from the fact that we are often using very limited computational resources, such as smartphones. As our research aims at application in robotic systems with such limitations, we measure the sampling rates on multiple platforms, including Android mobile devices. The time efficiency of our method is studied in section \ref{section:timeEfficiency}. Furthermore, when resources are scarce, e.g. when the number of particles is very low, reducing sampling variance can have a noticeable effect on particle filter performance. The analytic study of systematic sampling variance is very difficult \cite{douc05comparison, hol2006resampling}. Therefore we rely on numerical results and measure the sampling quality with the Cram{\'e}r-Von Mises goodness-of-fit statistic \cite{gentle03random} (see section \ref{subsection:cramer}). The sampling quality is studied in section \ref{section:quality}, and its effect in particle filtering is verified in section \ref{section:non-linear} using a standard non-linear time series. 

\section{Systematic Alias Sampling (SAS)}\label{section:sas}
\subsection{Alias Method}
Walker \cite{walker74new, walker77efficient} proposed a very clever method to generate unbiased random numbers from a discrete distribution. The method is based on the fact that the distribution can be expressed as a equiprobable mixture of $n$ two-point distributions. The mixture can be constructed in $O(n)$ by utilizing two working sets for greater and smaller-than-average elements, as shown by Kronmal and Peterson \cite{kronmal79alias}. After construction the structure can be stored in three arrays of length $n$: \textit{values}, \textit{aliasedValues}, and \textit{aliasProbabilities}. Sampling can then be done in constant time by first randomly determining the bin $i$ and then randomly choosing \textit{values}$(i)$ or \textit{aliasedValues}$(i)$ with biased coin based on \textit{aliasProbabilities}$(i)$ (see Algorithm \ref{alg:sample}). We use the terms \textit{lower bin} and \textit{upper bin} as synonyms for elements in \textit{values} and \textit{aliasedValues} respectively, as the alias table structure intuitively consists of upper and lower bins (see Figure \ref{fig:aliasTable}, middle). 

When approximating one-dimensional continuous distributions while using systematic sampling, it is beneficial to keep the values in spatial order. We try to achieve this in the upper and lower bins by using stacks as working sets when generating the alias table. Figure \ref{fig:aliasTable} illustrates the spatial ordering. Optimal table structure, also referred to as the alias table generation problem \cite{smith05analysis}, is out of the scope of this paper.

\begin{algorithm}[t!]
\scriptsize
\DontPrintSemicolon
\caption{Sampling with the alias method \;
\centerline{\textbf{sample}$(i, x)$}}
\label{alg:sample}
\KwIn{ \\
$i \in [0, n-1]$: random or otherwise produced integer \;
$x \in [0, 1[$: random or otherwise produced real number
}
\KwOut{
sample from the discrete distribution
}
\BlankLine
	\eIf{$x \leq$ \textit{aliasProbabilities}$(i)$}{
	\Return{ \textit{aliasedValues}$(i)$ }
	}
	{
	\Return{ \textit{values}$(i)$ }
}
\end{algorithm}

\subsection{Cram{\'e}r-Von Mises goodness-of-fit statistic} \label{subsection:cramer}
For measuring sampling quality we use the Cram{\'e}r-Von Mises goodness-of-fit statistic, as it is more powerful for detecting deviations in the tail of the distribution than e.g. the Kolmogorov-Smirnov test \cite{gentle03random, spinelli94cramer, stachniss2007analyzing}. For discrete distributions the statistic can be formulated as follows \cite{spinelli94cramer}:
\[
W^2 = n^{-1}\sum_{i=0}^{n}(F_k(i) - F(i))^2,
\]
where $F_k$ is the empirical distribution function (edf) based on the $k$ samples and F is the cumulative distribution function (cdf). Throughout the paper we present the results as the square root of the statistic ($W$).

\subsection{The basic idea}\label{subsection:idea}
The very basic idea of Systematic Alias Sampling (SAS) is to randomly generate $k$  equidistant points in interval $[0, n[$ that are then mapped on the alias table structure to generate the samples. This follows the idea of systematic resampling \cite{douc05comparison}: we first generate uniformly a random number $r_0$ from interval $[0, \frac{n}{k}[$ that systematically determines a set of real-valued points $\{r_0 + i \cdot  \frac{n}{k} \vert i = 0,...,k-1 \}$. The integer and fractional part of the points determine the bin and the choice between the upper and lower value respectively. This is illustrated in Figure \ref{fig:aliasTable} and the pseudo code is presented in Algorithm \ref{alg:systematicSample}. As sampling from the alias table is $O(1)$, the overall time complexity of SAS is $O(k)$. This combined with the fact that we avoid almost all random number generation, makes the method very fast in practice, as demonstrated in section \ref{section:timeEfficiency}. 

\subsection{Problems with divisibility and almost divisibility}\label{section:divisibility}

\begin{figure*} [ht]
 \centering
 \includegraphics[width = 1.0\textwidth]{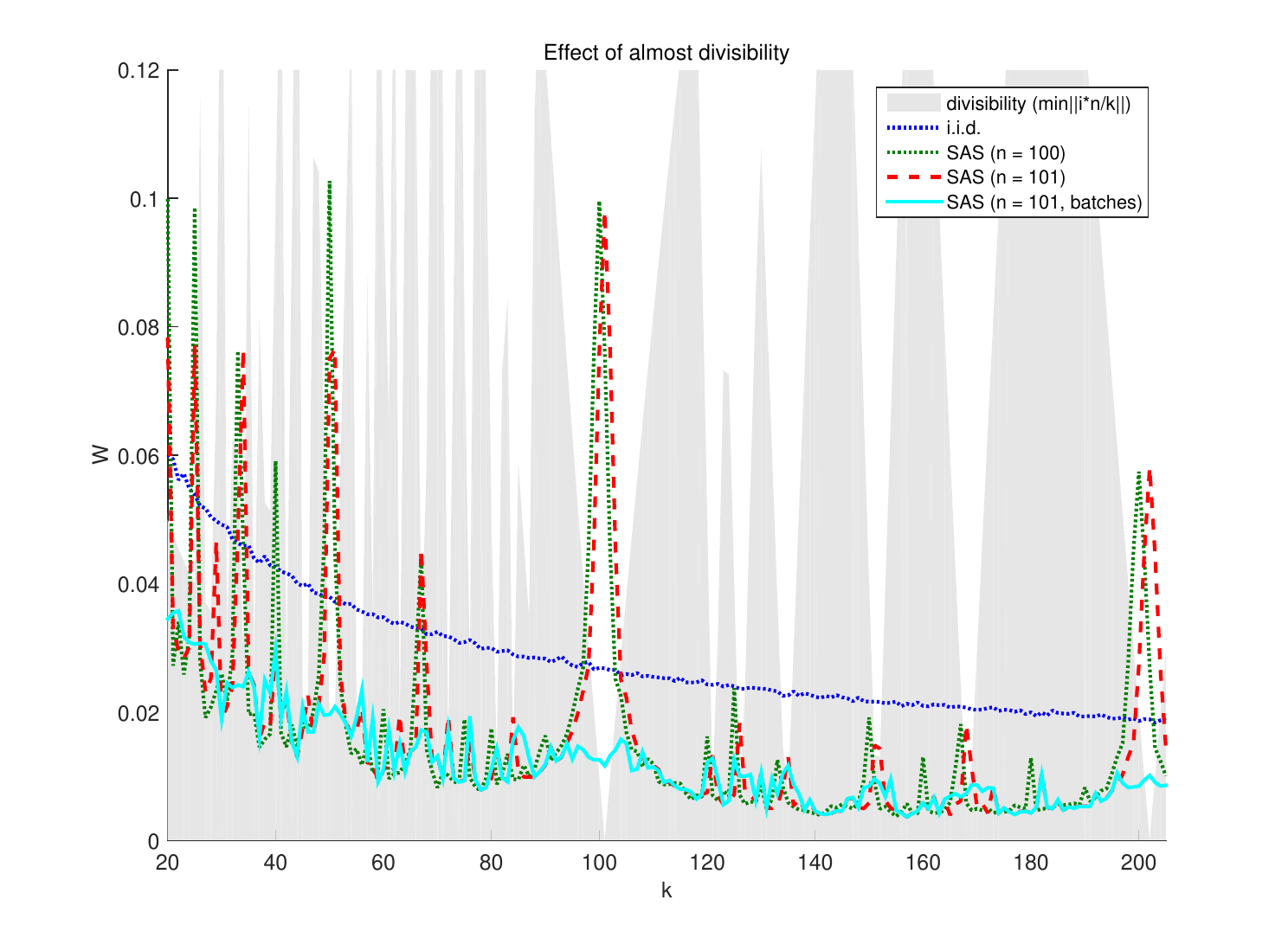}
 \caption{
 Divisibility problems caused by systematically sampling from the alias table structure. Issues with the 101-valued tailed distribution (Fig. \ref{fig:tailedDistribution}) can be clearly seen in the Cram{\'e}r-Von Mises statistic (section \ref{subsection:cramer}). Choosing bin size to be a prime number somewhat alleviates the problem. When we divide batches based on almost divisibility, SAS outperforms i.i.d. sampling regardless of sample count. The almost divisibility $\min\{\Vert \frac{i \cdot n}{k} \Vert\ | i\in\{1,4,5,6\}\}$ is visualized in the background. The results are averaged over 1000 runs.
 }
 \label{fig:divisibility}
\end{figure*}

However, this kind of systematic approach does not come without complications. The alias table structure indeed can cause an unwanted correlation resulting in spurious artefacts. For example, it can be easily seen that when $k$ divides $n$, the fractional parts of the sampling points are the same and all of the samples come from the same level of the alias table structure. This leads to upper or lower bins being overrepresented in the batch, causing serious malformation in the empirical distribution that is reflected clearly in the Cram{\'e}r-Von Mises statistic. The same is true to a lesser extent also when $k$ divides $2n$ or other multiplicatives of $n$. The phenomenon is illustrated in Figure \ref{fig:divisibility}.

\begin{figure} [ht]
 \centering
 \includegraphics[width = \columnwidth]{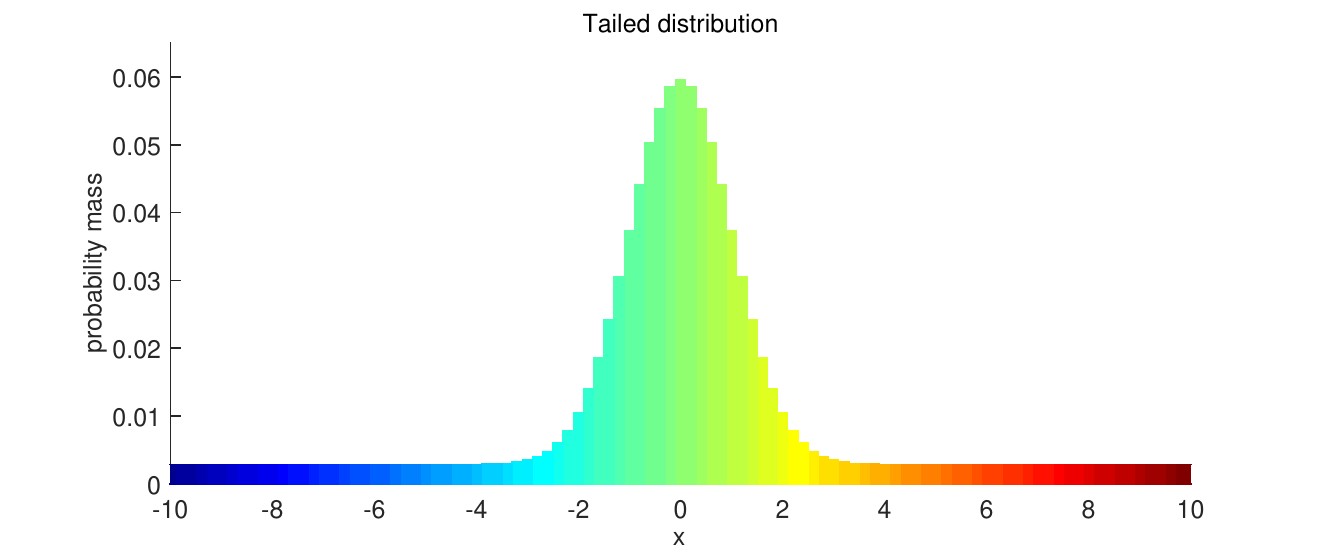}
 \caption{
 Tailed 101-valued distribution used in the experiments to make sampling artefacts more prominent. 
 }
 \label{fig:tailedDistribution}
\end{figure}   

To better reveal the artefacts we sample from a tailed distribution (Fig \ref{fig:tailedDistribution}). For this we chose a 101-valued mixture defined by the probability mass function (pmf) $\hat{f}(x)_{101} := c(\phi(x) + 0.02$) with support of [-10, 10], where $\phi$ is the probability density function (pdf) of the standard normal distribution and $c$ is the normalizing constant. Similar distributions are often used in robotics to cope with measurement outliers and to mitigate inaccuracies in proposal distributions \cite{thrun05probabilistic, thrun2001robust}.

The first step to alleviate this problem was to choose $n$ to be a prime number (the original distribution can always be modified by adding dummy values, e.g. as in the alias-urn method \cite{peterson82mixture}). While the problem is somewhat reduced (Fig. \ref{fig:divisibility}), it is clearly caused by \textit{almost divisibility} \cite{erdos83almost} as well:

\begin{definition}
Let $x$ be a real number, $ \{x\} = x - \lfloor x \rfloor $ its fractional part, and $ \Vert x \Vert = \min(\{x\}, 1-\{x\})$ denoting its distance from the nearest integer. We say that the positive real number $b$ is \textit{almost divisible} by the positive real number $a$ if $\Vert \frac{b}{a} \Vert$ is small. More exactly, we say that if $\varepsilon > 0$ and $\Vert \frac{b}{a} \Vert < \varepsilon$, then $b$ is \textit{$\varepsilon$-divisible} and $a$ is an \textit{$\varepsilon$-divisor} of $b$. In this case, we write $a\vert_\varepsilon b$.      
\end{definition}

Our simple solution to the divisibility problem is to analyse if $k$ almost divides $n$ or some of its multiplicatives. If this is the case and $k \geq k_{\min}$, we call the \textit{sampleSystematic} method (see Algorithm \ref{alg:systematicSample}) recursively in two batches and concatenate samples from \textit{sampleSystematic}($k-l$) and \textit{sampleSystematic}($l$) for the final result.

To find suitable values for the parameters we measured the sampling quality ($W$) with bin counts 101, 251, 503, and 1009 (prime numbers close to typical approximation granularities). The multiplicatives of $n$ included in the divisibility check were $i \cdot n$, where $i \in \{1,4,5,6\}$. This set was experimentally found to yield good results for its size. By analysing $\min\{\Vert \frac{i \cdot n}{k} \Vert\ | i\in\{1,4,5,6\}\}$, when $k = 1,\dots,2n$, we found that a reasonable value for $\varepsilon$ was $0.07$. Details about the found set for divisibility check and value of $\varepsilon$ are presented in section \ref{section:epsilon}.  The minimum batch size for recursion $k_{\min} := 15$ was chosen so that it does not $\varepsilon$-divide $n$. The recursive batch sizes were  determined as follows: if $k < 4k_{\min}$, we set $l = k_{\min}$; otherwise, we set $l = \lfloor 6/13 \cdot k \rfloor$. We shall note here that the method is not particularly sensitive to exact parameter values, and other choices were found to work as well.

The recursive approach yields significantly improved results, and effectively makes even the problematic sample counts produce more fit empirical distributions  than i.i.d. sampling. The divisibility problems and the effect of the recursive approach for the 101-valued tailed distribution (Fig. \ref{fig:tailedDistribution}) are illustrated in Figure \ref{fig:divisibility}. We acknowledge that the divisibility problem is not completely removed by this approach, but at least for the tested bin counts it offers a reasonable and quick practical remedy. When $k > n$, we suggest using systematic sampling, based on our experiments in sections \ref{section:quality}, \ref{section:timeEfficiency}, and \ref{section:non-linear}. We analyse the sampling quality further in section \ref{section:quality}.

\begin{figure}
 \centering
 \includegraphics[width = \columnwidth]{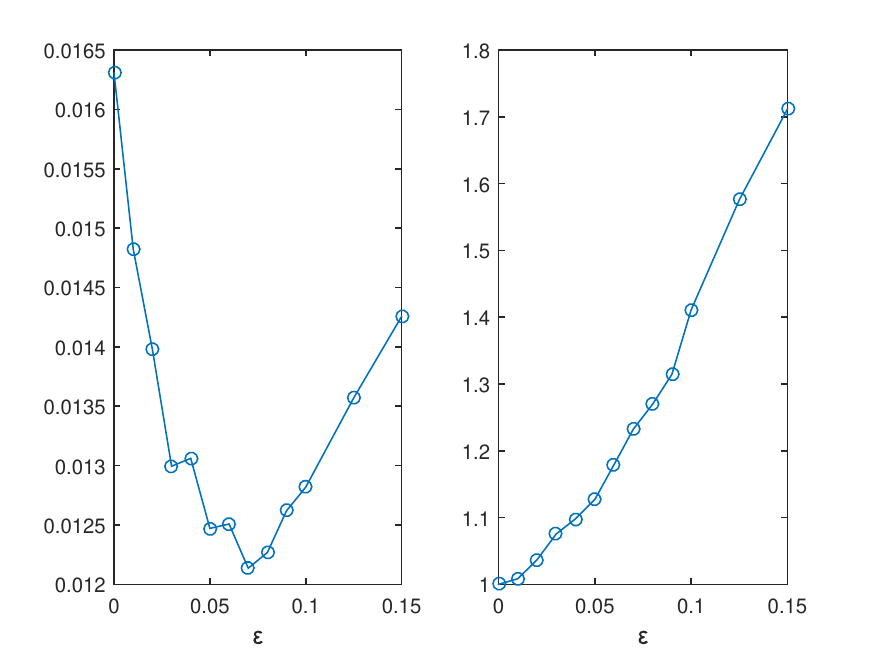}
 \caption{
 Left: Cram{\'e}r-Von Mises statistic for different values of $\varepsilon$. Right: Sampling time for different values of $\varepsilon$ relative to $\varepsilon = 0$. The results are averaged over sample counts $k \in [20,205]$ from the results for the tailed distribution, SAS $(n = 101)$ with batch division (see Figure \ref{fig:divisibility}).
 }
 \label{fig:epsilon}
\end{figure}

\subsection{The set for divisibility check and choice of $\varepsilon$} \label{section:epsilon}
The experimental integer set $S$ used in the divisibility check was found by analysing how well $\min\{\Vert \frac{i \cdot n}{k} \Vert\ | i\in S\}$ detected the peaks in the Cram{\'e}r-Von Mises statistic. As we have to check the divisibility against all of the chosen multiplicatives, we want to keep $S$ minimal. By analysing the problematic values of $k$, and experimenting with integers corresponding to multiplicatives of $n$ that are almost divisible by such $k$, we found a small but effective set $S := \{1,4,5,6\}$.  This choice detected virtually all of the most problematic intervals and yielded very good results for all of the experimented bin counts. Figure \ref{fig:divisibility} illustrates the correspondence of the problematic values of $k$ and the almost divisibility (gray background). With fixed $S$, the value of $\varepsilon$ acts as a threshold for peak detection. We want to keep $\varepsilon$ as small as possible for two reasons. The first reason is performance: avoiding multiple batch generation shows clearly in the running time, as illustrated in Figure \ref{fig:epsilon} (right). The second reason is that the more we resort on multiple batch generation, the closer we are to i.i.d. sampling. This is depicted in Figure \ref{fig:epsilon} (left) as an incline in the Cram{\'e}r-Von Mises statistic as $\varepsilon$ becomes larger than $0.07$. We note here that as $\varepsilon$ depends on $S$, its optimal value may vary by $S$. In our experiments we fix $S := \{1,4,5,6\}$ and subsequently set $\varepsilon := 0.07$.

\subsection{Systematic Alias-urn sampling (SAS-urn)} \label{section:urn}
Another way to tackle the divisibility problem is to construct the alias table as in the alias-urn method \cite{peterson82mixture} by artificially inflating the number of one-valued bins. The original intent of the method is to trade memory for performance by avoiding some of the random number generations as well as table accesses. However, in our case the main advantage is that by using more memory we can simplify the alias table structure and reduce the divisibility artefacts significantly, as the number of one-valued bins is inflated.

\subsection{Quasi-random sequences as seeds} \label{section:golden}
Quasi-random or low-discrepancy sequences are one popular way to generate evenly distributed values in unit cube \cite{niederreiter92random}. The numbers $x_i$ produced by a one-dimensional quasi-random sequence can be used to sample values from the alias table. This is done by first drawing a random number $u_0$ from U(0,1) and then generating the samples by calling \textit{sample}($\lfloor y_i \rfloor , y_i - \lfloor y_i \rfloor$), where $y_i := n \cdot (x_i + u_0) \pmod{n}$. Alternatively one can generate multiple batches without initializing the sequence in between just by using the next $k$  elements to generate the samples. Advantages of using quasi-random sequences are that the produced samples do not need to be shuffled, and any number of samples can be requested at a time without any initialization overhead.  However, at least the implementations of Sobol and Halton sequences in ACM 3.4 are too slow for our use, and produce very poor results when $n$ is close to the power of two minus one, i.e. $2^m - 1$ (fractional parts seem to correlate strongly). On the contrary the Golden Ratio sequence \cite{schretter12golden} is very fast, easy-to-implement, and does not seem to be affected by such correlation problems. Therefore we chose the Golden Ratio sequence to be included in our experiments. The Golden Ratio sequence is defined as follows:
\[
x_i = x_{i-1} + \phi \pmod{1},
\]
where $\phi = \frac{1 + \sqrt{5}}{2} \approx 1.61803...$ is the golden ratio and $x_0$ is drawn from U(0,1). 

\subsection{Implementation} \label{section:implementation}
Algorithm \ref{alg:systematicSample} outlines our implementation of the SAS algorithm. As can be seen from the pseudo code, the algorithm requires a very low number of very simple operations. After initialization (lines 1-5), to generate one sample, we only use one integer comparison, one cast from floating point to integer, one integer and two floating point additions, and (inside Alg. \ref{alg:sample} \textit{sample} method) two array accesses and one floating point comparison. 

We traverse the alias table structure backwards (lines 3 and 9) in order to prevent index going over array bounds due to numerical inaccuracies in the floating point addition (line 9). The rationale behind this is that if we get a similar rounding error in the zero-end, we end up calling \textit{sample}($0, -\epsilon$) that maps correctly to the same value as \textit{sample}($0, 0$).

In our actual initialization implementation we also applied the divisibility-check (see section \ref{section:divisibility}) and provide an output array with a starting index as parameters in order to avoid concatenation and array creation overhead in the recursion. In the pseudo code these are omitted for code clarity.
\begin{algorithm}[t!]
\scriptsize
\DontPrintSemicolon
\LinesNumbered
\caption{Systematic Alias Sampling (SAS)\;
\centerline{\textbf{sampleSystematic}$(k)$}}
\BlankLine
\label{alg:systematicSample}
\KwIn{ 
$k$: number of samples
}
\KwOut{
$k$ samples from the discrete distribution
}
\BlankLine
$step \gets n/k$\;
$r_0 \sim U[0, step[$ \;
$x \gets n - r_0$\;
samples $\gets$ new array of $k$ elements \;
$i \gets 0$\;
\While{$i < k$}{
$j \gets \lfloor x \rfloor$ \;
samples($i$) $\gets \textbf{sample}(j, x - j)$\tcc{Alg. \ref{alg:sample}}
$x \gets x - step$ \;
$i \gets i + 1$ \;
}
\Return{samples}
\end{algorithm} 
 
\section{Sampling quality}\label{section:quality}
 
Convergence speed of a random sequence is well-studied in the quasi-random number literature \cite{niederreiter92random, morokoff94quasi}. Sequences obtained by i.i.d. sampling converge almost surely due to the law of large numbers However, the sampling variance in i.i.d. sampling increases the convergence time. In our context we can measure the sampling quality by analysing the empirical distribution function defined by the generated $k$ samples. For this we use the Cram{\'e}r-Von Mises statistic (section \ref{subsection:cramer}), that measures how well the empirical distribution function fits the given discrete distribution. It is particularly suitable for our use as it is sensitive to tail malformations. For the same reason in this experiment we use the tailed distribution defined in Section \ref{section:divisibility} (Fig. \ref{fig:tailedDistribution}). This is to make tail sampling artefacts and their effect on Cram{\'e}r-Von Mises statistic more prominent.

We measured the Cram{\'e}r-Von Mises statistic for empirical distributions produced by four different methods: Systematic sampling (Systematic), Systematic Alias Sampling (SAS), Systematic Alias Sampling that uses Golden Ratio sequence as seeds (SAS-golden), and Systematic Alias-urn Sampling (SAS-urn). We also computed i.i.d. sampling results for comparison. The Systematic sampling \cite{douc05comparison} generates the samples by computing a sequence of $k$ equidistant points $\{r_0 + i \cdot  \frac{n}{k} \vert i = 0,...,k-1 \}$, where $r_0$ is a uniform random number from interval $[0, \frac{n}{k}[$. These numbers are then mapped to the cdf of the distribution to generate the samples (e.g. by using binary search). SAS is the method described in Section \ref{section:sas} and uses the divisibility check to divide the batches to alleviate the effects of almost divisibility. SAS-golden uses Golden Ratio sequence as seeds as described in \ref{section:golden} and uses no divisibility check. SAS-urn (see \ref{section:urn}) uses a radically inflated bin count ($11n$) and utilizes the same divisibility check as SAS.

\begin{figure} [ht]
 \centering
 \includegraphics[width = \columnwidth]{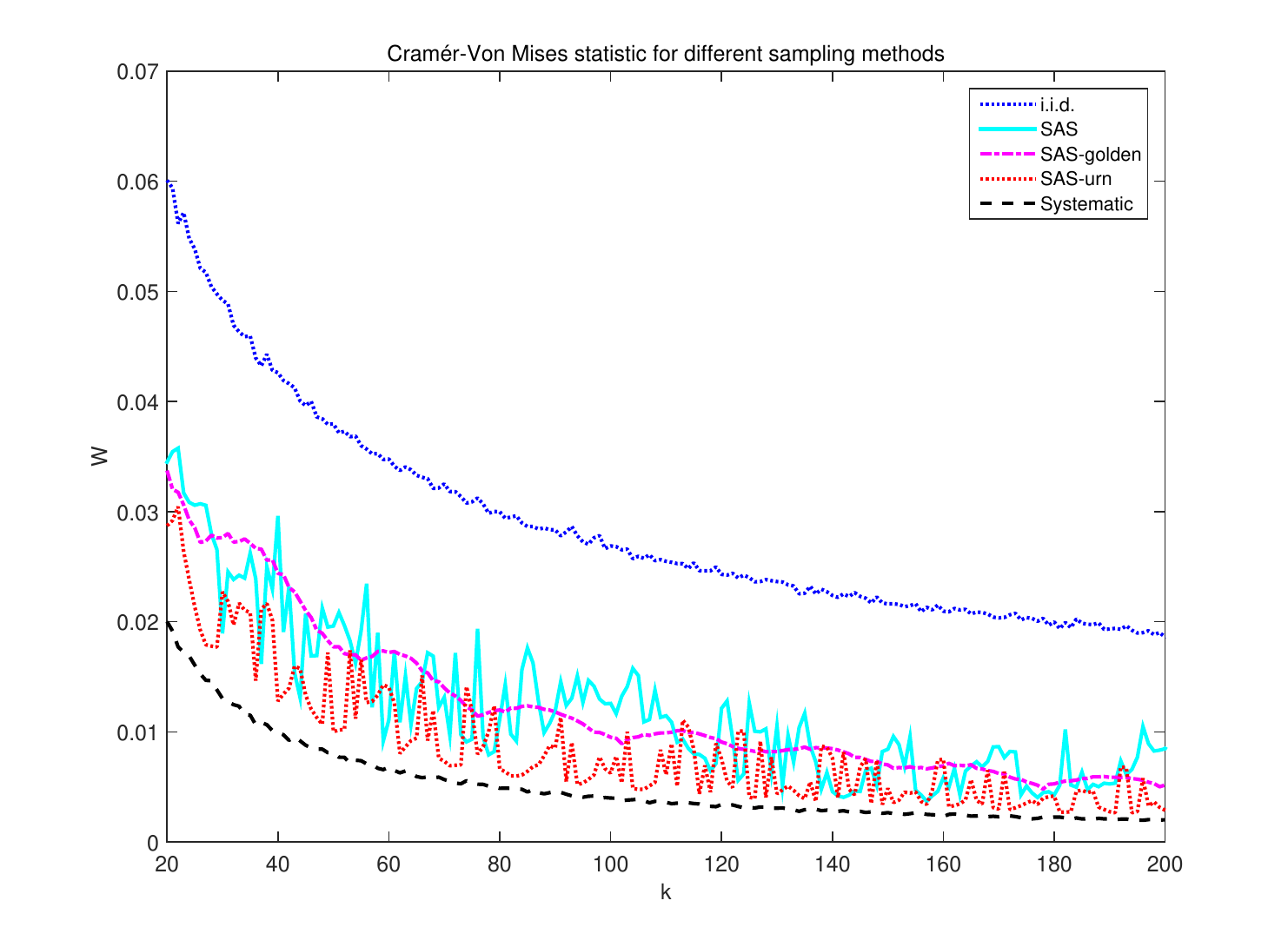}
 \caption{Cram{\'e}r-Von Mises statistic for different sampling methods in the tailed distribution experiment (Section \ref{section:divisibility} and Fig. \ref{fig:tailedDistribution}). The results are averaged over 1000 runs. }
 \label{fig:cramerQuality}
\end{figure}

Figure \ref{fig:cramerQuality} illustrates the Cram{\'e}r-Von Mises statistic for different methods. Not surprisingly the systematic sampling has the best goodness-of-fit, followed in order by SAS-urn, SAS, and SAS-golden. All systematic methods outperform i.i.d. sampling clearly. The variation due to almost divisibility can be clearly seen in the SAS and SAS-urn plots, whereas SAS-golden produces a more consistent fit. Although SAS-golden performance does not vary so much based on $k$, its overall goodness-of-fit is slightly worse due to not utilizing the spatial correlation as effectively as SAS and SAS-urn. Table \ref{table:quality} summarizes the Cram{\'e}r-Von Mises statistic relative to i.i.d. sampling for different sampling methods and bin counts. The relative values are computed as an average over both $k = 1,\dots,2n$ and $1000$ runs. We additionally experimented SAS-golden without the uniform random number initialization between batches, but the results were identical to the pre-batch initialization.

\begin{table}[ht]
\scriptsize
\centering
\caption{Cram{\'e}r--von Mises statistic (W) relative to i.i.d.\ sampling for different sampling methods.}
\label{table:quality}
\begin{tabular}{ l | c | c | c | c }
  \hline
  $n$   & Systematic & SAS  & SAS-golden & SAS-urn \\
  \hline      
  101   & 0.20 & 0.42 & 0.43 & 0.31 \\
  251   & 0.13 & 0.34 & 0.44 & 0.25 \\
  503   & 0.09 & 0.29 & 0.35 & 0.17 \\
  1009  & 0.06 & 0.27 & 0.35 & 0.13 \\
  \hline  
\end{tabular}

\vspace{1ex}
\parbox{0.9\linewidth}{\small \textit{Note:} The relative values are computed as an average over both $k = 1,\dots,2n$ and 1000 runs.}
\end{table}


\section{Time efficiency}\label{section:timeEfficiency}
In order to measure the sampling performance of our method we generated 100 million samples in different batch sizes. To make the timing comparisons more robust, we measured the sampling rates on four platforms, including two Android 4.4 devices with different virtual machines: Nexus 5 (with ART) and 2012 Nexus 7 (with Dalvik). The other two platforms were a Windows 7 desktop computer and a Windows 8.1 laptop, both running 64-bit Java SE 7 (Update 51). All experiments were run on a single core. Processor information is given in Table \ref{table:samplingRates}. Choosing Java allows easier transition to and comparable running times on the Android mobile devices. The actual code was written in Scala programming language that compiles to Java bytecode, because it offers specialized generic classes for Java primitive types for primitive-like performance \cite{dragos10compiling}. This kind of microbenchmarking is known to be notoriously difficult to get right. For that reason we additionally verified the results with the Google Caliper microbenchmarking framework \cite{google13caliper}.

For a relative comparison baseline we chose the ACM NormalDistribution \textit{sample(k)} method (ACM-ND) that we compared against different methods sampling from a discrete approximation of the standard normal distribution ($n = 1009$). We chose six methods for the comparison: 

\begin{itemize}
\item \textit{Binary Search.} The standard binary search implementation from ACM \cite{apache33}, used as in section \ref{section:quality} to generate i.i.d. samples by mapping random numbers to the cdf. 
\item \textit{Alias.} The alias sampling \cite{walker74new, walker77efficient}, that samples i.i.d. samples from the distribution (non-systematic).
\item \textit{SAS}. Systematic Alias Sampling detailed in section \ref{section:implementation} with parameters set identical to the experiment in section \ref{section:quality}.
\item \textit{SAS-golden}. Systematic Alias Sampling that uses the Golden Ratio sequence as seeds, as described in section \ref{section:golden}.
\item \textit{Systematic Binary.} The first version of systematic sampling generates the $k$ evenly spaced points in $[0,1[$, as described in section \ref{subsection:idea} and \cite{douc05comparison}. It then  uses the binary search to find the corresponding values in the cdf resulting in $O(k\log(n))$ complexity.
\item \textit{Systematic.} The second version of systematic sampling is similar to the implementation in \cite{arulampalam02tutorial} that iterates through the elements in the cdf, resulting in $O(n+k)$ complexity.
\end{itemize}

\begin{figure} [ht]
 \centering
 \includegraphics[width = \columnwidth]{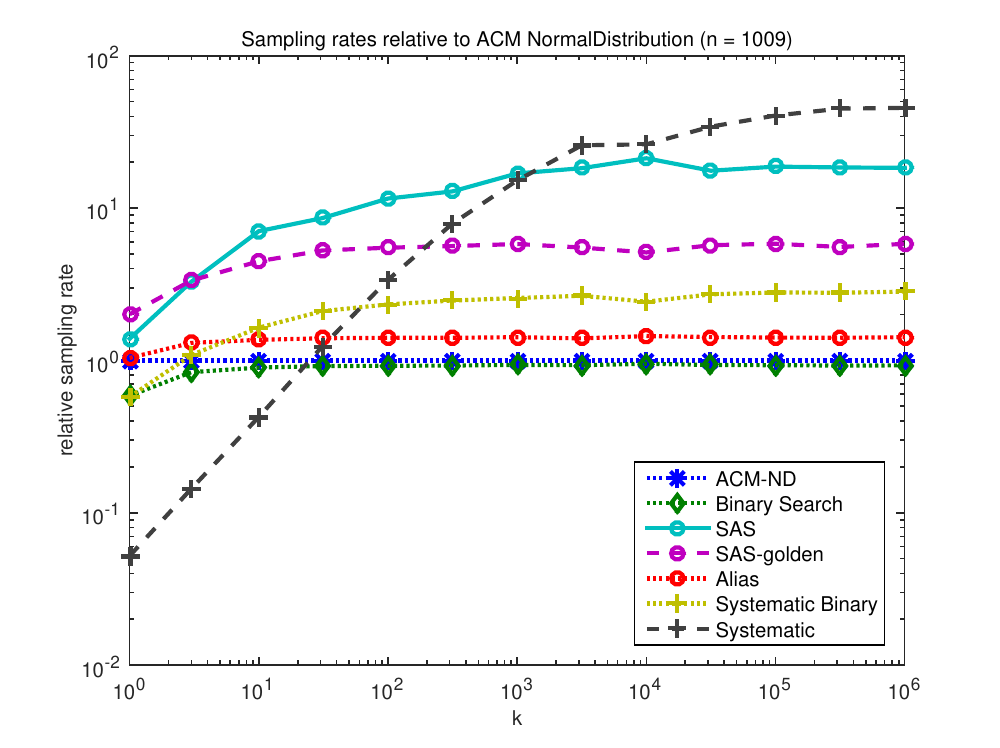}
 \caption{Sampling rates relative to Apache Commons Math NormalDistribution sample(k) method (ACM-ND). SAS ($n = 1009$) produces samples up to 20 times faster, and clearly outperforms systematic sampling when $k < n$. The results are in log-log scale.}
 \label{fig:samplingRates1009}
\end{figure}

For fair comparison to ACM-ND, all sampling methods include the result array allocation, although it has no noticeable effect on performance. All random numbers in the experiments were created using the Well19937c \textit{nextDouble()} method \cite{apache33}. The relative sampling rates for the methods are visualized in Figure \ref{fig:samplingRates1009}.

\begin{figure} [ht]
 \centering
 \includegraphics[width = \columnwidth]{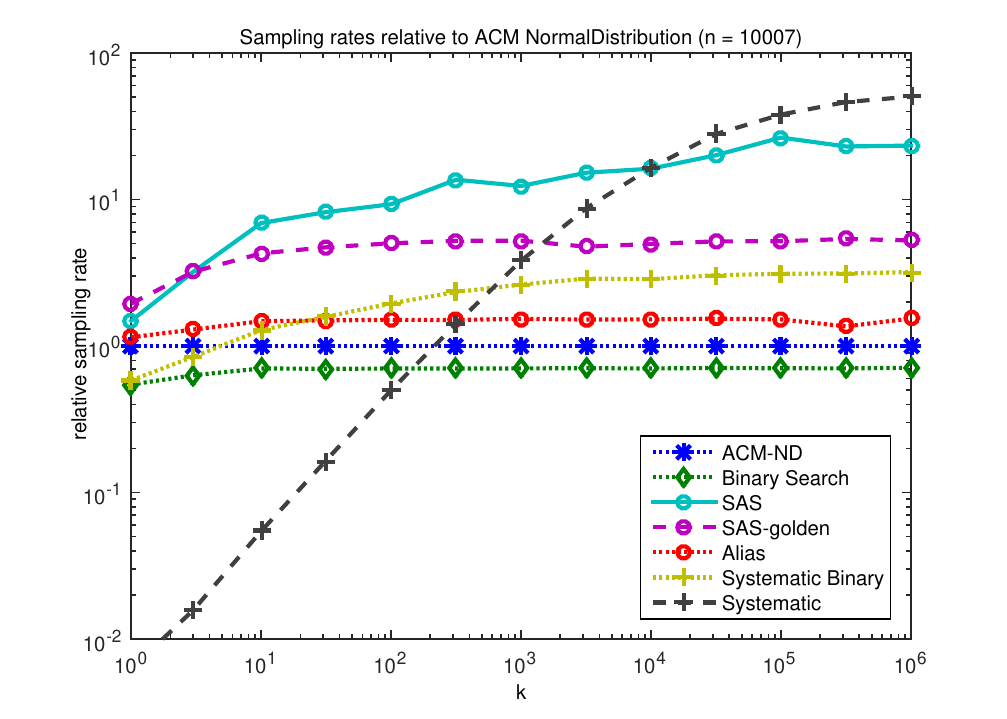}
 \caption{Sampling rates relative to Apache Commons Math NormalDistribution sample(k) method (ACM-ND). SAS ($n = 10007$) produces samples up to 20 times faster, and clearly outperforms systematic sampling when $k < n$. The results are in log-log scale.}
 \label{fig:samplingRates10007}
\end{figure}

The relative performance of SAS rapidly improves as $k$ grows from $1$ to $1000$, and SAS generates samples approximately $15$ times faster than ACM-ND when $k \geq 1000$. SAS-golden achieves a more modest rate ($\approx 6\times$), still outperforming systematic binary ($\approx 3\times$) and the alias sampling ($\approx2\times$). Unlike systematic sampling, the performance of SAS is independent of $n$ (approximation granularity), as can be verified by comparing Figures \ref{fig:samplingRates1009} and \ref{fig:samplingRates10007}. When $k < n$, SAS clearly outperforms all other methods, and only after $k$ becomes larger than $n$, systematic sampling begins to slightly dominate SAS. SAS-urn performed similarly to SAS (omitted from the pictures for clarity).

The relative sampling rates presented here are from the desktop platform. However, the results on all platforms are very similar to the desktop environment. Average sampling rates of $100$ million samples generated in batches of $100$ on different devices along with relative sampling rates are presented in Table \ref{table:samplingRates}. Because performance depends on e.g. compiler optimizations, cache locality, and computer platform, these kinds of microbenchmarking results should be interpreted cautiously. Nonetheless, because the results are similar on numerous test platforms, we expect that the relative performance will be approximately the same independent of environment.

\begin{table}[t!]
\centering
\scriptsize
\caption{Sampling rates on different platforms (millions/s) and performance relative to ACM NormalDistribution (ACM-ND).}
\label{table:samplingRates}
\begin{tabularx}{\columnwidth}{@{} l *{3}{>{\centering\arraybackslash}X} @{}}
\toprule
Device & ACM-ND & SAS & Speedup $\times$ \\
\midrule
Desktop, i5-3570@3.40\,GHz   & 14.57  & 168.3 & 11.6 \\
Laptop, i7-3517U@1.90\,GHz   & 11.69  & 168.3 & 14.4 \\
Nexus 5, Krait 400@2.26\,GHz & 0.778  & 10.08 & 13.0 \\
Nexus 7, Cortex A9@1.2\,GHz  & 0.186  & 1.83  & 9.8  \\
\bottomrule
\end{tabularx}

\vspace{0.4ex}
\parbox{\columnwidth}{\footnotesize\textit{Note:} Averages are over 100 million samples generated in batches of 100.}
\end{table}

\section{Performance on non-linear time series}\label{section:non-linear}
The non-linear time series is a well-studied problem often used to demonstrate how particle filters outperform Kalman filters in non-linear problems \cite{arulampalam02tutorial, carpenter99improved}. Here we use it to achieve two goals. First, we want to verify that reducing sampling variance when sampling from the transition model has an actual effect on particle filter performance. Similar analysis has been done for Lattice particle filters \cite{ormoneit2001lattice}. Second, we want to measure our method's performance on a particle filtering system that on some level resembles real use cases, in which e.g. cache performance can be considerably different to microbenchmarking.

As the error function we use the weighted average error of the particles before resampling, as we think it describes particle filter error better than a single estimate error. When computing the error, we ignore the sign as in \cite{hol2004resampling}. Otherwise we keep the parameters as in \cite{arulampalam02tutorial}. For the transition model sampling we use a discrete 1009-point approximation of the standard normal distribution, with support of $6.7$ standard deviations (approximately the limit for the Box-Muller transform as well). The performance is measured when using systematic samples generated by SAS, SAS-urn (as in section \ref{section:quality}), \textit{systematic sampling} (binary search version), and i.i.d. samples generated by the ACM NormalDistribution \textit{sample(k)} method. Because the systematic methods produce spatially correlated samples, the samples have to be shuffled before being used to propagate the particles. For this we simply use a precomputed random permutation.

\begin{figure}[ht]
 \centering
 \includegraphics[width = \columnwidth]{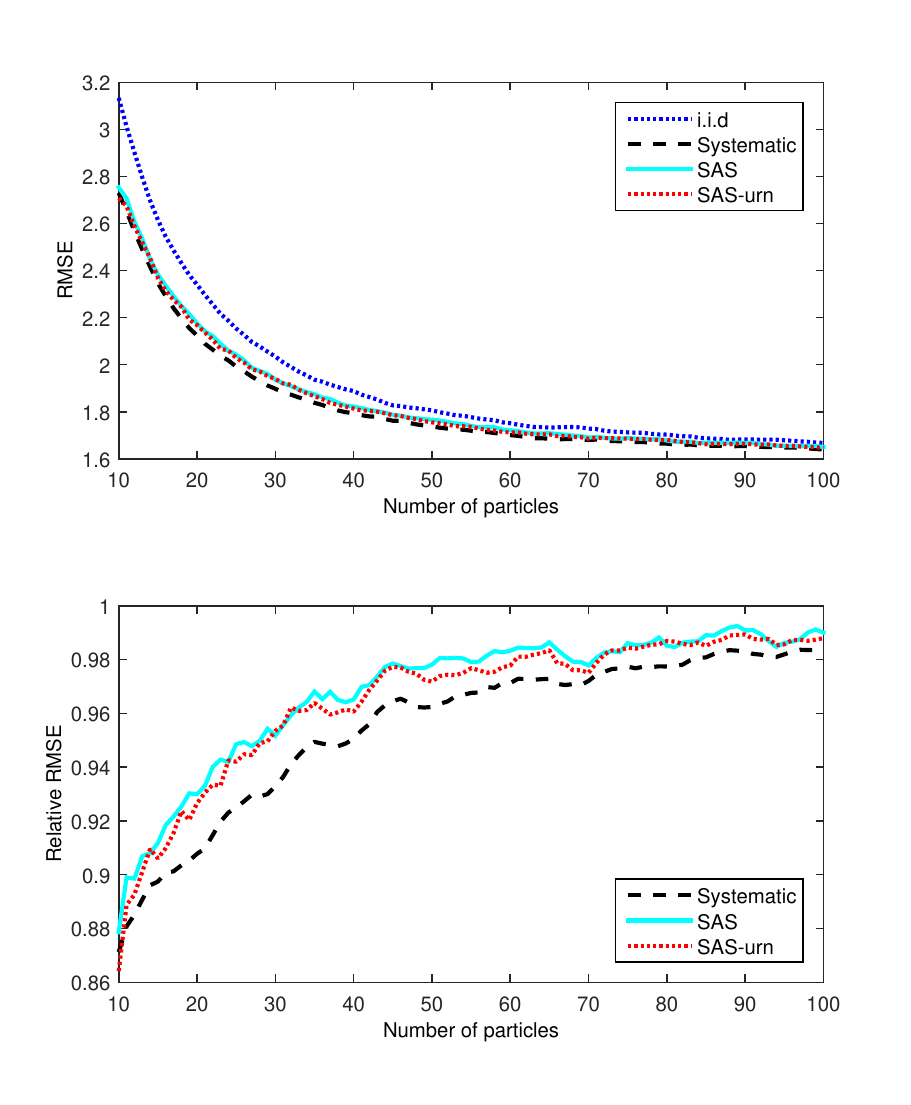}
 \caption{ Top: The RMSE for different particle counts used in the non-linear time series experiment. The i.i.d. sampling is outperformed by the systematic sampling methods especially when the number of particles is low. Bottom: RMSE relative to i.i.d. sampling reveals differences between systematic sampling methods more clearly. Systematic sampling outperforms the two other methods. SAS-urn is marginally better than SAS. The results are averaged over $1000$ instances of the non-linear time series.}
 \label{fig:nonlinearBenchmark}
\end{figure}

\begin{figure}[ht]
 \centering
 \includegraphics[width = \columnwidth]{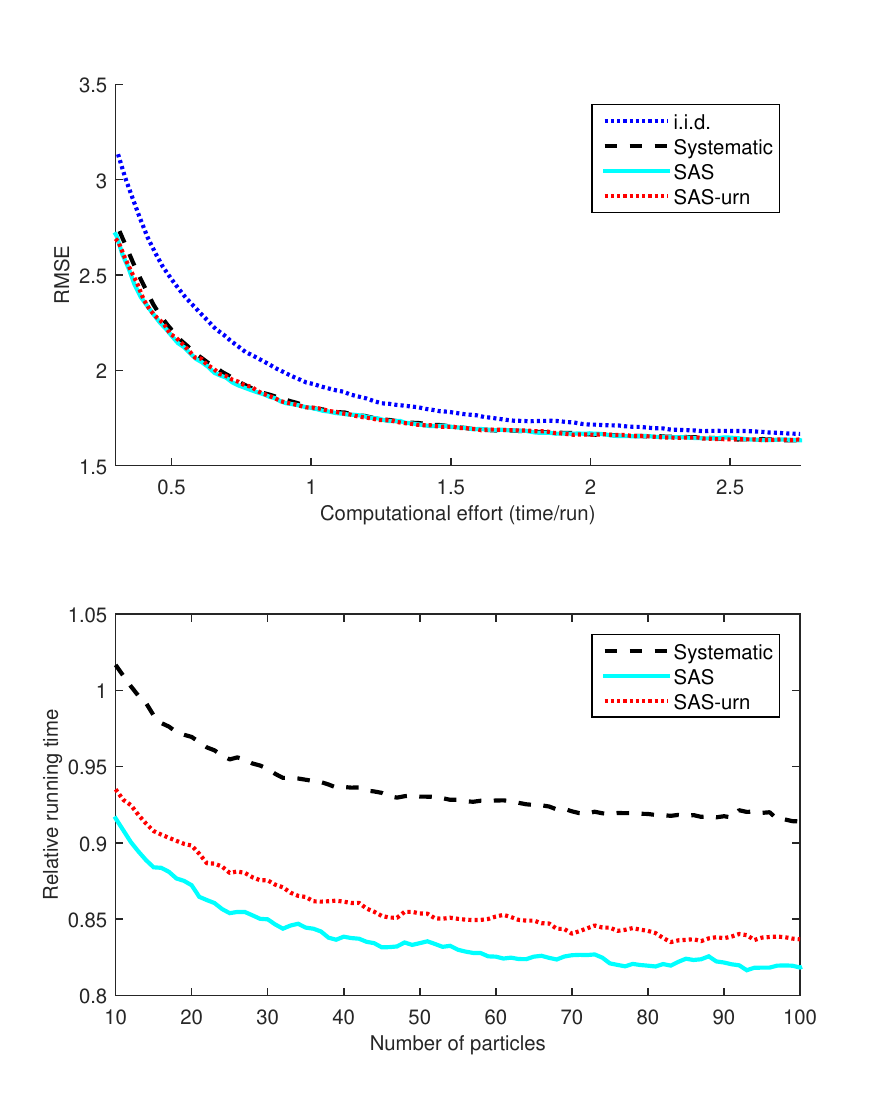}
 \caption{Top: RMSE w.r.t computational effort. The systematic methods yield almost identical results outperforming the ACM-ND \textit{sample(k)} method (i.i.d.). Bottom: Running times relative to the ACM-ND \textit{sample(k)} method. The results are averaged over $1000$ instances of the non-linear time series.}
 \label{fig:nonlinearBenchmarkTime}
\end{figure}

We measure the performance with particle counts ranging from $10$ to $100$. The results are averaged over $1000$ instances of the non-linear time series. Figure \ref{fig:nonlinearBenchmark} illustrates how the systematic methods outperform the i.i.d. sampling in the transition model, systematic sampling having the lowest RMSE. The results are in line with the ones presented in  \cite{ormoneit2001lattice}. However, as expected, the benefit due to smaller sampling variance soon diminishes as the number of particles grows. In addition we measure the running times and compare them to i.i.d. sampling. Because particle filter time complexity is linear in the particle count, we can easily compare the RMSE to computational effort (time/run). Figure \ref{fig:nonlinearBenchmarkTime} illustrates the running times relative to ACM NormalDistribution \textit{sample(k)} method as well as RMSE compared to computational effort. Although the \textit{systematic sampling} has a clear edge when measuring RMSE w.r.t particle count (Fig. \ref{fig:nonlinearBenchmark}), the advantage disappears when measuring w.r.t. computational effort. In this case the the performance of the systematic methods is almost identical (Fig. \ref{fig:nonlinearBenchmarkTime}). Despite a considerably larger memory footprint, SAS-urn does not produce a noticeable improvement over SAS. The reduced time performance is probably due to increased cache misses caused by larger arrays. In spite of the better measured sampling quality of SAS-urn (section \ref{section:quality}), the authors consider SAS preferable in practice due to its smaller memory footprint.

\section{Discussion and future considerations}
The method presented here is meant to be a general plug-in replacement for sampling random values from a desired discrete distribution. When more details of the problem are known, the parameters (e.g. $k_{\min}$) can be optimized for used sample and bin counts based on empirical performance. For example choosing batch sizes without divisibility problems would lead to some performance improvement due to avoiding the divisibility checks.

The presented method could be extended to multidimensional cases, because multivariate distributions can also be approximated by discrete distributions. The simplest way to do this would be to use a version that is seeded with quasi-random numbers (e.g. SAS-golden), as spatial order of the values is not trivial in the multidimensional case. Also cleverly ordering the bins (spatial indexing) could produce good results in practice. If the components are independent, one can sample from the marginal distributions, though some kind of shuffling is needed if SAS is used.

In typical particle filtering problems the number of particles is much higher than in the toy example studied in section \ref{section:non-linear}. However, the problems are also much harder, and a very small subpopulation or cluster of particles may be responsible for covering (distinct) part of the state space, making the low particle count performance relevant. For these kinds of situations it might be very beneficial to use cluster-wise systematic sampling, i.e. generate low-variance samples for one cluster at a time, thus reducing sampling variance inside the clusters. 

As the method is very lightweight and effective, the authors could imagine it useful also in computer game development as a convenient tool to generate random samples for e.g. particle effects. If implemented in fixed-point arithmetic, it could probably be utilized also in embedded systems lacking floating point unit.

\section{Conclusions}
In this paper we have presented a convenient and effective method to generate batches of $k$ random samples from an arbitrary discrete distribution. By avoiding random number generation and other costly operations we achieve sampling rates that are up to 20 times faster than in common mathematics libraries. Furthermore, the produced empirical distribution clearly outperforms i.i.d. sampling in quality in terms of the Cram{\'e}r-Von Mises statistic. We demonstrated that this leads to improved performance when used for sampling in a particle filter transition model especially when the number of particles is low.

\bibliography{mSLAM}

\end{document}